\documentclass[aps,prl,twocolumn,groupedaddress]{revtex4}
\usepackage{epsfig}

\begin{document}

\title{Direct Calculation of the Critical Effective Potential}

\author{H. Ballhausen}
\email{physics@ballhausen.com}
\affiliation{Institute for Theoretical Physics, Heidelberg University\\Philosophenweg 16, 69120 Heidelberg, Germany}

\begin{abstract}
\noindent
The critical effective potential is the nonperturbative part of the effective action
at a phase transition. It equals the scale invariant effective average potential and
can be calculated from the renormalization group flow of the effective average action.
In some cases this requires only the solution of an ordinary differential equation 
without actually simulating the renormalization group flow. Here the Ising model is 
examined beyond leading order and with full field dependent effective potential.
\end{abstract}

\maketitle

\bigskip
\noindent
\textbf{Introduction}

\bigskip
\noindent
The effective potential in statistical physics and quantum field theory \cite{Lit1,Lit2} includes fluctuations
of all wavelenghts as opposed to the microscopic potential. It is therefore relevant for the
computation of macroscopic properties.

\bigskip
\noindent
The usual way to access the effective potential is to start with the potential given by the 
microscopic theory and follow the renormalization group flow to the infrared \cite{Lit2a}. 
As the potential then depends on field variables and the scale, this requires solving a 
partial differential equation.

\bigskip
\noindent
However, an especially interesting case is a second order phase transition where the correlation
length diverges. The system becomes therefore scale invariant and the renormalization group
flow ends up in a fix point, the effective critical potential. Its calculation requires only
the solution of an ordinary differential equation \cite{LitA,LitB,LitC,LitD,LitE,LitF,LitG,LitH}.

\bigskip
\noindent
~

\bigskip
\noindent
\textbf{Effective average action}

\bigskip
\noindent
The effective average action ( see e.g. \cite{Lit3} and references therein ) $\Gamma_k$ interpolates between the microscopical or 
classical action $S=\Gamma_\infty$ and the effective action $\Gamma=\Gamma_0$. By construction it includes only fluctuations 
with momenta larger than $k$ and this transition is described by an exact renormalization group flow equation:

\bigskip $ \partial_k \Gamma_k = \textrm{Tr} \Big( (\Gamma_k^{(2)} + R_k)^{-1} \partial_k R_k \Big) $

\bigskip
\noindent
Here $\Gamma_k^{(2)}$ is the two point function and $R_k$ is an arbitrary momentum cutoff with the properties $R_k\to0$ for $k\to0$,
$R_k\to\infty$ for $k\to\infty$ and $R_k(q^2)>0$ for $q^2\to0$.

\bigskip
\noindent
We want to consider a particular interesting and common model, the three dimensional Ising model. The $O(1)$ symmetry of 
this theory can be exploited for a lowest order derivative expansion of the effective average action:

\bigskip $ \Gamma_k = \int d^dx \Big( U_k(\rho) + Z_k \frac{1}{2} \partial^\mu \varphi(x) \partial_\mu \varphi(x) \Big) $

\smallskip $ \rho=\frac{1}{2} \varphi(x)^2 ~~~~~~~~~~ \varphi \in I\!\!R^3 \to I\!\!R $

\bigskip
\noindent
in terms of the most general form of the nonperturbative effective average potential $U_k(\rho)$ and a field independent but 
scale dependent wave function renormalization $Z_k$.

\bigskip
\noindent
When inserted into the flow equation for the effective average action this ansatz gives the flow of the effective average potential.

\bigskip $ \partial_k U_k = \frac{1}{2} \int \frac{d^dq}{(2\pi)^d} \frac{ \partial_k R_k(q^2)}{ U_k^\prime + 2 \rho U_k^{\prime\prime} + q^2 Z_k + R_k(q^2) } $

\bigskip
\noindent
As the potential is only determined up to a constant, 
it is convenient to consider its first derivative instead. The corresponding flow equation is given by straightforward differentiation:
 
\bigskip $ \partial_k U_k^\prime = \frac{1}{2} \int \frac{d^dq}{(2\pi)^d} \frac{ (- 3 U_k^{\prime\prime} - 2 \rho U_k^{\prime\prime\prime}) \partial_k R_k(q^2)}{(U_k^\prime + 2 \rho U_k^{\prime\prime} + q^2 Z_k + R_k(q^2))^2} $

\bigskip
\noindent
Switching to dimensionless quantities $u^\prime=U_k^\prime/(Z_k k^2)$ and choosing a linear cutoff function $R_k=Z_k(k^2-q^2)\Theta(k^2-q^2)$ \cite{Lit4,Lit4a} 
the flow equation becomes explicitely scale invariant:

\bigskip $ \partial_t u^\prime(x) = (-2+\eta) u^\prime + (d-2+\eta) x u^{\prime\prime} $

\smallskip $ ~~~~~~~~~\; - 4 v_d \frac{d+2-\eta}{d(d+2)} \, \frac{3u^{\prime\prime}+2x u^{\prime\prime\prime}}{(1+u^\prime+2x u^{\prime\prime})^2} $

\bigskip
\noindent
Here $t=\textrm{Ln}(k)$, $x = Z_k k^{2-d} \rho$, $v_d^{-1}=2^{d+1}\pi^{d/2}\Gamma(d/2)$ 
and $\eta$ denotes the anomalous dimension $\eta=\partial_t\textrm{Ln}Z_k$.

\bigskip
\noindent
The latter one can either be taken self consistently from 

\bigskip $ \eta = \frac{8v_d}{d} \kappa \frac{(3\lambda+2\kappa u^{(3)})^2}{(1+2\kappa \lambda)^4} $

\bigskip
\noindent
( where $\kappa(t)$ denotes the running minimum ( $u^\prime(\kappa)=0$ ) and $\lambda=u^{\prime\prime}(\kappa)$, 
$u^{(3)}=u^{\prime\prime\prime}(\kappa)$ ) or from the quite accurate approximation

\bigskip $ \eta = d^{-d} $

\bigskip
\noindent
which will be used in this paper.

\bigskip
\noindent
For comparison with literature we also use a sharp cutoff, where in leading order

\bigskip $ \partial_t u^\prime = -2 u^\prime + (d-2) x u^{\prime\prime} - 2 v_d \, \frac{3u^{\prime\prime}+2x u^{\prime\prime\prime}}{1+u^\prime+2x u^{\prime\prime}} $

\newpage
\noindent
\textbf{Scaling solutions}

\bigskip
\noindent
The scale invariance of some system at a second order phase transition manifests itself in a fix point
in the renormalization group flow of the dimensionless, scale invariant flow equations. We are thus 
interested in scaling solutions $u^\prime$ such that

\bigskip $ 0 = \partial_t u^\prime = ac u^\prime + bc x u^{\prime\prime} + c \frac{3u^{\prime\prime}+2 x u^{\prime\prime\prime}}{(1+u^\prime+2 x u^{\prime\prime})^2} $

\bigskip
\noindent
where

\bigskip $ a = \frac{d(d+2)}{4v_d} \, \frac{2-\eta}{d+2-\eta} ~~~~~~~~ b = \frac{d(d+2)}{4v_d} \, \frac{-d+2-\eta}{d+2-\eta} $

\bigskip
\noindent
This can be rewritten canonically in terms of first order ordinary differential equations for $u^\prime(x)$ and $u^{\prime\prime}(x)$:

\bigskip $ \partial_x u^\prime = u^{\prime\prime} $

\smallskip $ \partial_x u^{\prime\prime} = \frac{-1}{2x} \Big( (au^\prime+bxu^{\prime\prime})(1+u^\prime+2xu^{\prime\prime})^2+3u^{\prime\prime} \Big) $

\bigskip
\noindent
Given initial values $u^\prime_0$, $u^{\prime\prime}_0$ at some point $x_0$ the differential 
equations yield a local solution $u^\prime$, $u^{\prime\prime}$ in a vicinity around $x_0$. 
The space of such solutions has dimension two, its elements being uniquely defined by 
$u^\prime_0$, $u^{\prime\prime}_0$.

\bigskip
\noindent
However, we are interested only in physical solutions which are continuations of the local
solutions to the whole positive real axis. In other words, we must find a point $x_0\geq0$ 
and initial values $u^\prime_0$, $u^{\prime\prime}_0$ such that $u^\prime$ is a smooth 
function for all $x\geq0$.

\bigskip
\noindent
Such a search is nontrivial, though, even numerically. As soon as the space of local solutions
has a dimension larger than one, global solutions can be missed even if one is already close
to the correct initial values. Fortunately, by using the constraint of requiring a global 
solution we are able to eliminate one of the two integration constants:

\bigskip
\noindent
If $ \partial_t u^\prime = 0 ~\forall x $, this holds true especially for $x_0=0$:

\bigskip $ 0 = \partial_t u^\prime \mid_{x_0} \, = ac u^\prime_0 +c \frac{3u^{\prime\prime}_0}{(1+u^\prime_0)^2} $

\bigskip
\noindent
And hence 

\bigskip $ u^{\prime\prime}_0 = - \frac{a}{3} u^\prime_0 (1+u^\prime_0)^2 $

\bigskip
\noindent
In the analogous calculation for the sharp cutoff we find:

\bigskip $ 0 = \partial_t u^\prime = ac u^\prime + bc x u^{\prime\prime} + c \frac{3u^{\prime\prime}+2 x u^{\prime\prime\prime}}{1+u^\prime+2 x u^{\prime\prime}} $

\bigskip
\noindent
where $ a = v_d^{-1}$, $b = v_d^{-1}(1-d/2) $ such that

\bigskip $ \partial_x u^\prime = u^{\prime\prime} $

\smallskip $ \partial_x u^{\prime\prime} = \frac{-1}{2x} \Big( (au^\prime+bxu^{\prime\prime})(1+u^\prime+2xu^{\prime\prime})+3u^{\prime\prime} \Big) $

\bigskip $ u^{\prime\prime}_0 = - \frac{a}{3} u^\prime_0 (1+u^\prime_0) $

\newpage
\noindent
Furthermore, we have additional constraints on $u^\prime_0$: the physically interesting 
$u^\prime$ has a single zero and is positive for large $x$. Thus $u^\prime_0<0$. On the 
other hand, the denominator $1+u^\prime+2x u^{\prime\prime}$ must be nonzero for all $x$.
Therefore $u^\prime_0>-1$. Now the global solution can be found be varying only $u^\prime_0$, 
in the range $-1<u^\prime_0<0$, until the above requirements are met. This can be done 
numerically efficiently.

\bigskip
\noindent
For any initial $u^\prime_0$ the solution $u^\prime$ extends to some $x_{max}$. For too small
$u^\prime_0$ the flow diverges to positive infinity. On the other hand, for large initial 
values the flow crosses zero a second time. These two cases are separated by a critical 
initial value which leads to the physical solution we are interested in. The plot of 
$x_{max}$ as a function of $u^\prime_0$ is then sharply peaked around this critical value:

\bigskip
\noindent
\begin{center}
\epsfig{file=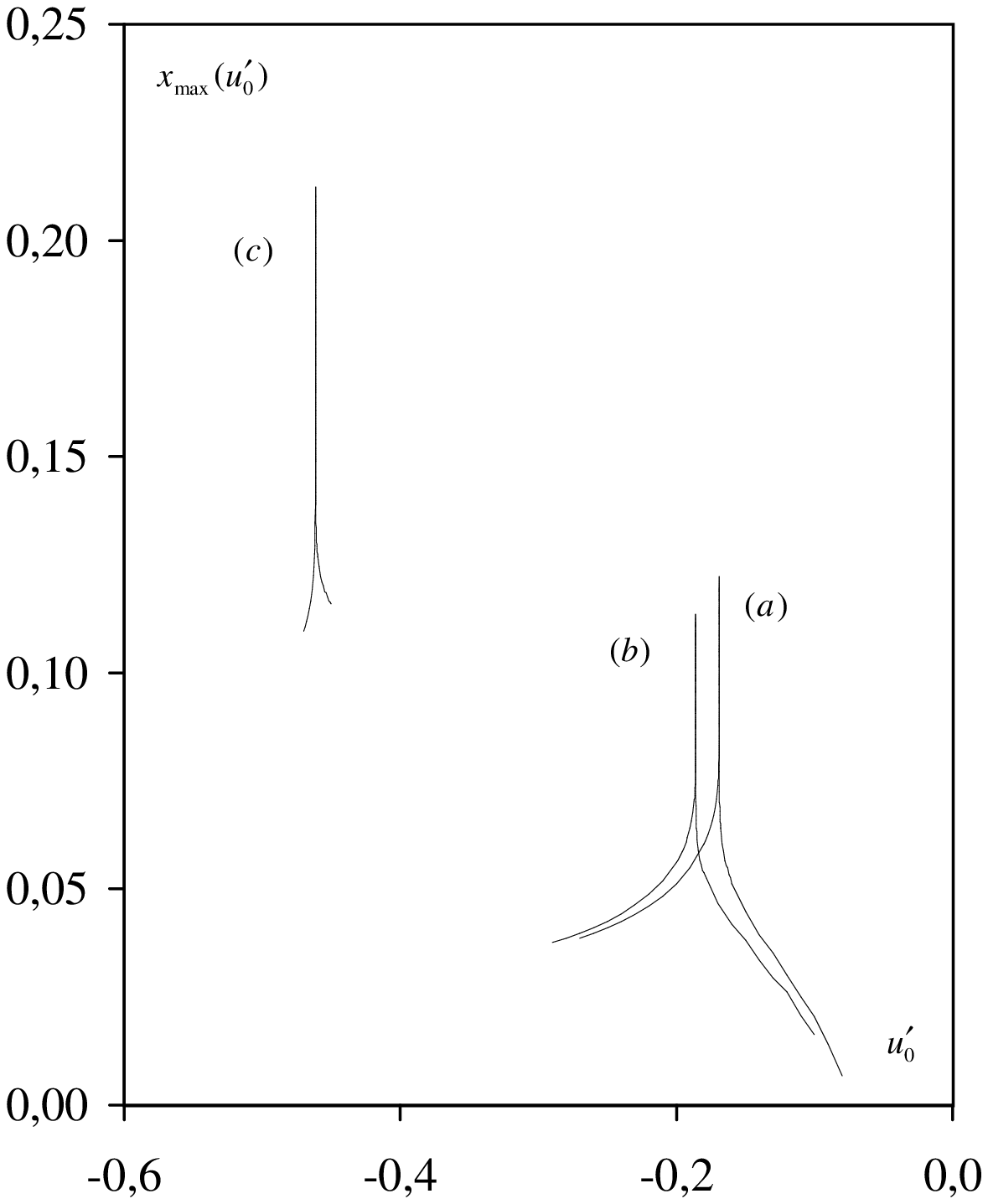,width=8.5cm}
\small{FIG. I:\\$x_{max}(u^\prime_0)$ for the linear cutoff (a),\\$\eta=0$ (b) and the sharp cutoff (c).}
\end{center}

\bigskip
\noindent
In this way we find:

\bigskip
\noindent
\begin{tabular}{ll}
$~~~$linear cutoff, $\eta=1/27$ (a)$~~~$&$u^\prime_0$=-0.16902863438...$~~~$\\
$~~~$linear cutoff, $\eta=0$ (b)     $~~~$&$u^\prime_0$=-0.18606424944...$~~~$\\
$~~~$sharp cutoff, $\eta=0$ (c)      $~~~$&$u^\prime_0$=-0.46153372007...$~~~$\\
\end{tabular}

\bigskip
\noindent
In perfect agreement with

\bigskip
\noindent
\begin{tabular}{ll}
$~~~$reference \cite{LitH} ( $\eta=0$ )     $~~~~~~~~\;$&$\lambda_{1*}$=-0.1860642...$~~~$\\
$~~~$reference \cite{LitB} ( $\eta=0$ )     $~~~~~~~~\;$&$\sigma$=-0.46153372...$~~~$\\
\end{tabular}

\newpage
\noindent
\textbf{Effective scaling potentials}

\bigskip
\noindent
We can now plot the solutions $u^\prime$ for the above initial values:

\bigskip
\noindent
\begin{center}
\epsfig{file=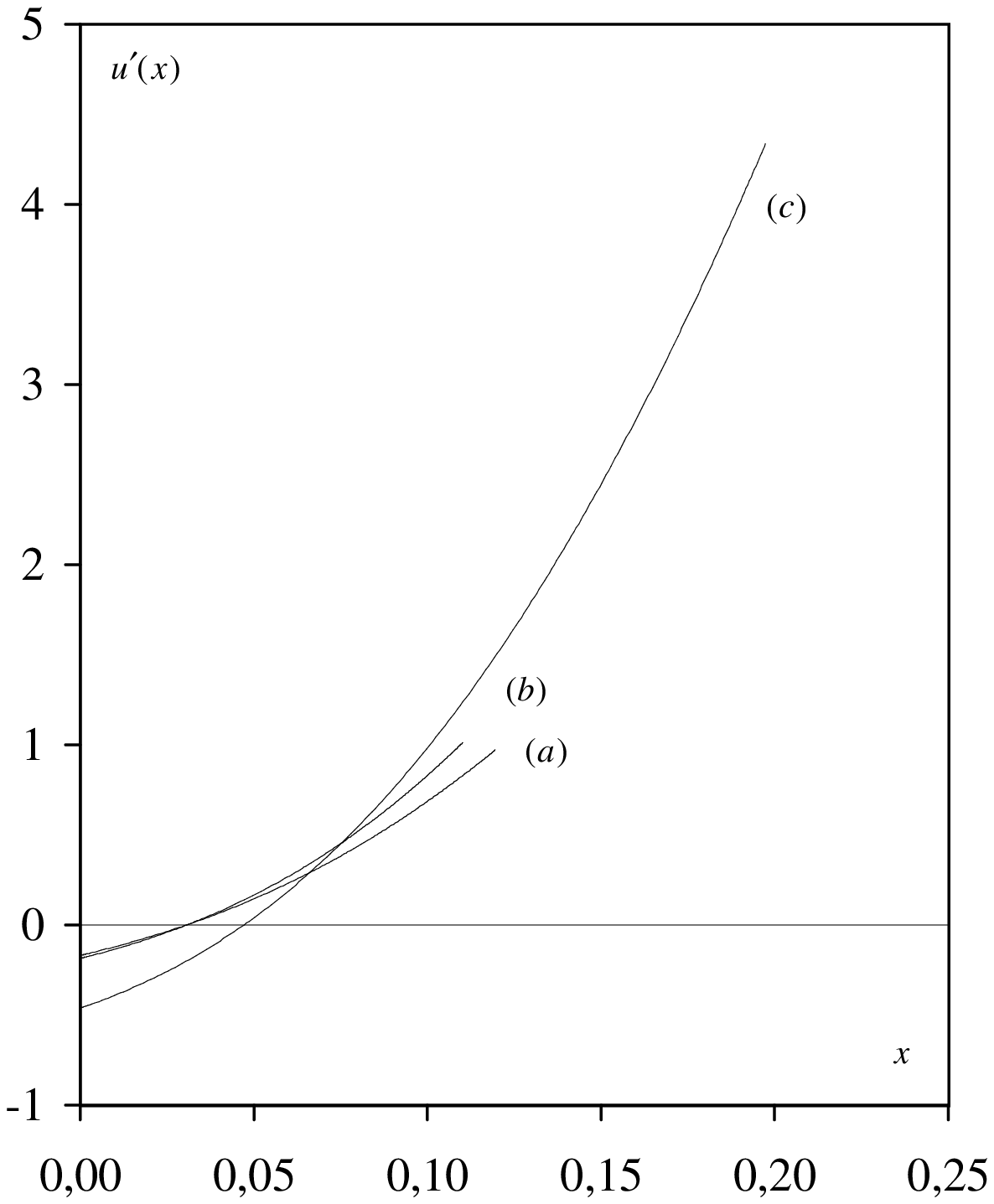,width=8.5cm}
\small{FIG. II:\\$u^\prime(x)$ for the linear cutoff (a),\\$\eta=0$ (b) and the sharp cutoff (c).\\Compare plot (b) to fig. 2 in \cite{LitH}.}
\end{center}

\bigskip
\noindent
Due to the numerically limited accuracy of the initial value and the integration we are able 
to track these solutions only up to $x\approx0.12$ in case of the linear cutoff and 
$x\approx0.2$ in case of the sharp cutoff.

\bigskip
\noindent
The second quantity to be read off from the critical effective potential is its zero
$ \kappa $ where $u^\prime(\kappa)=0$:

\bigskip
\noindent
\begin{tabular}{ll}
$~~~$linear cutoff, $\eta=1/27$ (a)  $~~~$&$\kappa$=0.03060106819...$~~~$\\
$~~~$linear cutoff, $\eta=0$ (b)     $~~~$&$\kappa$=0.03064764922...$~~~$\\
$~~~$sharp cutoff, $\eta=0$ (c)      $~~~$&$\kappa$=0.04711134650...$~~~$\\
\end{tabular}

\bigskip
\noindent
The value $\kappa$=0.03060... is in good agreement with
the fix point value $\kappa^*$=0.03053... from the dynamical method. 

\bigskip
\noindent
As one can see from the different solutions for the linear and the sharp cutoff, the
effective potential is not cutoff independent. Interestingly however, the location of 
the minimum hardly depends on the anomalous dimension in case of the linear cutoff. 
Moreover, it is also only weakly dependent on the initial value $u^\prime_0$.

\newpage
\noindent
\textbf{The critical exponent $\delta$}

\bigskip
\noindent
That the effective scaling potential in fact describes critical phenomena
can be seen from the power law \mbox{$u^\prime \sim x^{(\delta-1)/2}$}
for large $x$.

\bigskip
\noindent
The critical exponents $\delta$ and $\eta$ are related by

\bigskip $\delta=\frac{d+2-\eta}{d-2+\eta}$

\bigskip
\noindent
Hence we expect

\bigskip $ \lim \limits_{x\to\infty} 1+2 \frac{\partial \textrm{ln}(u^\prime)}{\partial \textrm{ln}(x)} = 5 $

\bigskip
\noindent
in case $\eta=0$ and

\bigskip $ \lim \limits_{x\to\infty} 1+2 \frac{\partial \textrm{ln}(u^\prime)}{\partial \textrm{ln}(x)} = \frac{67}{14} $

\bigskip
\noindent
in case $\eta=1/27$.

\bigskip
\noindent
These asymptotic values are plotted as horizontal lines:

\bigskip
\noindent
\begin{center}
\epsfig{file=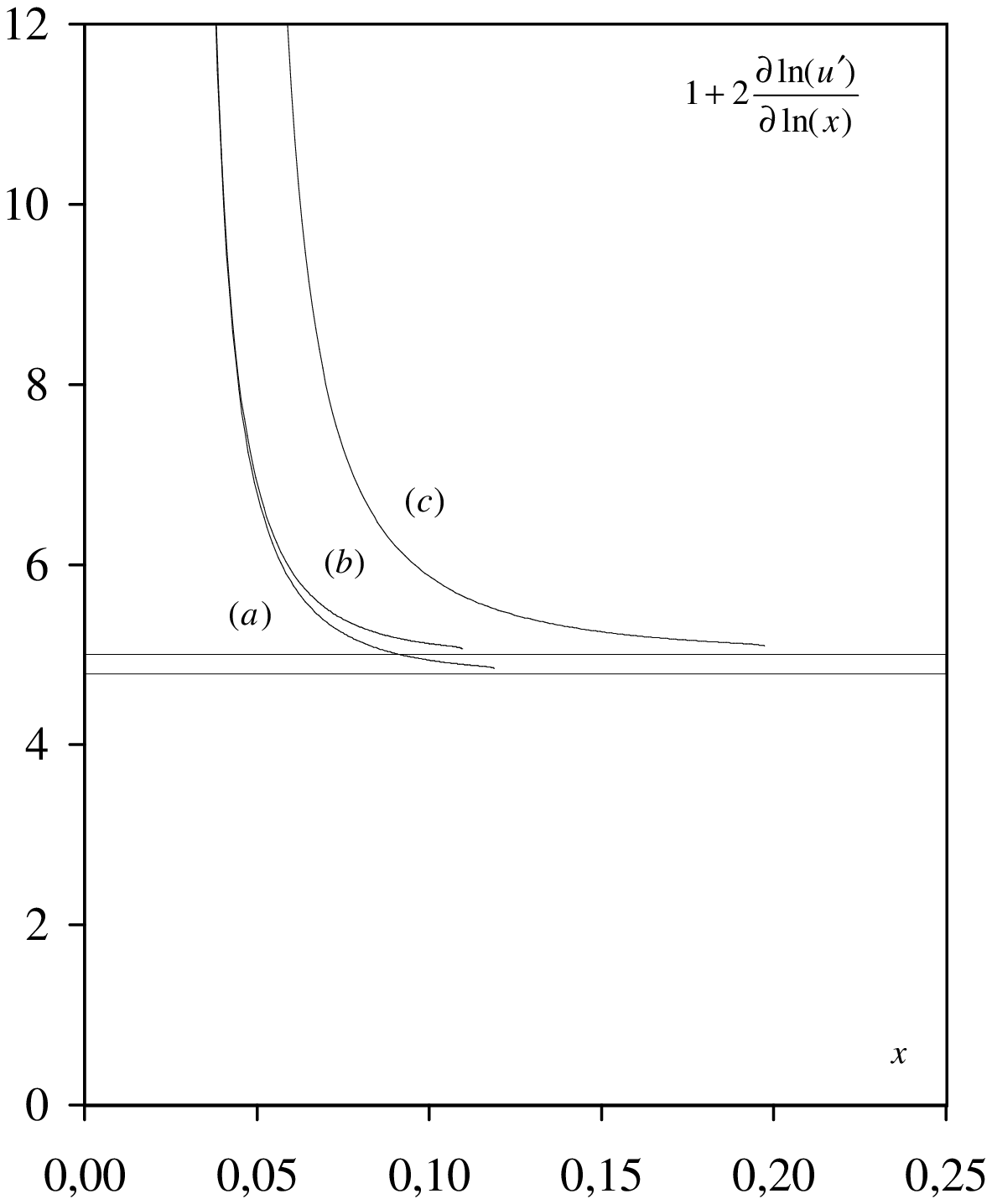,width=8.5cm}
\small{FIG. III:\\$1+2\partial \textrm{ln} u^\prime / \partial \textrm{ln} x$ 
for the linear cutoff (a),\\$\eta=0$ (b) and the sharp cutoff (c)\\
approaching the critical exponent $\delta$.}
\end{center}

\bigskip
\noindent
The asymptotic behaviour is as expected in all three cases.
However, if one wanted to calculate $\delta$ from the potential
its precision would suffer from the limited range of integration.

\newpage
\textbf{Comparison to the dynamical method}

\bigskip
\noindent
In order to assess the advantage in computing complexity of the method presented, we also take a
look at the usual dynamical method \cite{Lit2a,Lit5}. In this case one starts with an initial potential 
$u_\Lambda^\prime=\lambda_\Lambda(x-\kappa_\Lambda)$ at some ultraviolet scale $\Lambda$
with an initial expectation value of the fields $\kappa_\Lambda$ and some quartic coupling constant
$\lambda_\Lambda$.

\bigskip
\noindent
With this initial potential the partial differential equation for $\partial_t u^\prime(x)$
is computed on a discretized grid. The renormalization group flow for $t\to-\infty$ then either leads 
to the symmetric phase ( $\kappa \to 0$ ) or to the phase with spontaneous symmetry breaking ( $\kappa\to\infty$ ).

\bigskip
\noindent
These phases are separated by a second order phase transition. Correspondingly, there is a critical
$\kappa_\Lambda=\kappa_{cr}$. Fine tuning $\kappa_\Lambda$ this value can be found. The renormalization group
flow near the phase transition will then come close to the fix point and the scaling potential can be read out.

\bigskip
\noindent
The fine tuning of $\kappa_\Lambda$ is now replaced by fine tuning $f_0$. In effect, the solution of a
partial differential equation is replaced by the solution of an ordinary differential equation. This is
much simpler and may reduce computing time by several orders of magnitude.

\bigskip
\noindent
~

\bigskip
\noindent
\textbf{Conclusion and outlook}

\bigskip
\noindent
The method allows the calculation of the effective scaling potential with
very little effort. The integrable range seems to depend only on numerical precision and
could maybe be improved by the use of high accuracy floating point libraries.

\bigskip
\noindent
The effective scaling potential could also be used as an initial value in a dynamical 
simulation, allowing the determination of critical exponents dependend on small deviations from the critical
point, while still overcoming the time consuming necessity for a dynamical fix point search.

\bigskip
\noindent
~

\bigskip
\noindent
~

\end{document}